\documentclass[onecolumn,nofootinbib,preprintnumbers,superscriptaddress,longbibliography,10pt]{revtex4}

\usepackage{subfigure}
\usepackage{amssymb}
\usepackage{graphics}
\usepackage{graphicx}
\usepackage{amsmath}
\usepackage{amsfonts}
\usepackage{bm}

\newlength{\extraspace}
\newlength{\extraspaces}
\newcommand{\be}{\begin{equation}
\addtolength{\abovedisplayskip}{\extraspaces}
\addtolength{\belowdisplayskip}{\extraspaces}
\addtolength{\abovedisplayshortskip}{\extraspace}
\addtolength{\belowdisplayshortskip}{\extraspace}}
\newcommand{\ee}{\end{equation}}

\newcommand{\ba}{\begin{eqnarray}
\addtolength{\abovedisplayskip}{\extraspaces}
\addtolength{\belowdisplayskip}{\extraspaces}
\addtolength{\abovedisplayshortskip}{\extraspace}
\addtolength{\belowdisplayshortskip}{\extraspace}}
\newcommand{\ea}{\end{eqnarray}}

\usepackage{color,soul}
\usepackage[colorlinks,citecolor=blue,urlcolor=blue,linkcolor=blue]{hyperref}
\newcommand{\nonu}{\nonumber \\[.5mm]}

\begin{document}

\begin{center}
{\bf General form of the function $f(\mathbb{Q})$ using cylindrically static spacetime }
\end{center}
\begin{center}
{  Gamal Nashed$^{1,2}$  }\\
{ \it \small $^1$Center for Theoretical Physics, British University of Egypt,
Sherouk City 11837, Egypt\\
$^2$Center for Space Research, North-West University, Potchefstroom 2520, South Africa}
\end{center}
\hspace{2cm} \hspace{2cm}
\\
\\
\\
\\
\\
We find an exact static solution in four dimensions to the  field equations  of the $f(\mathbb{Q})$ gravity by using a cylindrically static spacetime with two different ansatz, $\nu(r)$ and $\mu(r)$. This solution is derived without imposing any conditions on   $f(\mathbb{Q})$. The black hole solution involves four constants: $c_1$, $c_2$, $c_3$, and $c_4$. Among these, $c_1$ is linked to the cosmological constant, $c_2$ to the black hole's mass, while $c_3$ and $c_4$ are responsible for the deviation of the solution from the linear form of $f(\mathbb{Q})$.  We demonstrate how the analytical function  $f(\mathbb{Q})$ relies on  $c_3$. When $c_3$ is zero, $f(\mathbb{Q})$ becomes a constant function, leading to the non-metricity case.  We investigate the singularity of this solution and show that the Kretschmann invariant has a much milder singularity compared to the non-metricity case.  We produce a black hole that rotates  with non-vanishing  values of $\mathbb{Q}$ and $f(\mathbb{Q})$ by using a coordinate transformation. Then, we analyze the laws of thermodynamics to determine the physical characteristics of this black hole solution and demonstrate that it is locally thermodynamically stable.
\section{Introduction}\label{S1}

Einstein's theory of relativity, called general relativity (GR), explains how gravity works by saying that spacetime curves. It's like a fabric that bends under the weight of objects. GR uses a type of math called Riemannian geometry to describe this curved spacetime. It's a way to understand how things move and interact under gravity. In metric-affine geometry, there are three main concepts: curvature, torsion, and non-metricity. They're all connected. When we look at gravity theories, like general relativity (GR), these concepts play a big role. For example, in GR, we don't consider torsion and non-metricity. But in other theories, like teleparallel equivalent to GR (TEGR) or symmetric teleparallel equivalent to GR (STEGR), different combinations of these features are important.

While GR  can explain many things, like how Mercury's path bends and how light bends near the Sun, it doesn't fully explain how the universe changes over time. Recent discoveries, like confirming black holes exist, show that GR works in many cases. But when we look at the universe's overall evolution, GR falls short. Observations of things like supernovas, Baryonic Acoustic Oscillations, the Cosmic Microwave Background, and large-scale structure, and experiments like WMAP all suggest that the universe is expanding faster over time, and GR doesn't fully account for this.   Dark energy is a special kind of energy that causes the universe to expand quickly in standard cosmology or the $\Lambda$-Cold-Dark-Matter ($\Lambda$-CDM) model. It's like a force that pushes everything apart, making the universe grow faster. Scientists sometimes change GR to better understand the mysterious parts of the universe, like dark energy and dark matter. They start by rethinking how things work in GR. This is the first thing they do when they want to change it. $f(R)$ gravity is a modified theory that considers the action as a function of the Ricci scalar $R$. This theory has been widely studied \cite{Nojiri:2003ft, Nojiri:2007as,Nashed:2018piz,Nashed:2020mnp}. More recently, a theory called $f(\mathbb{Q})$ has emerged, which is designed to understand the universe in the context of STEGR \cite{BeltranJimenez:2017tkd}. This theory has been revisited in various areas such as cosmology \cite{BeltranJimenez:2019tme,Paliathanasis:2023ngs,Paliathanasis:2023nkb,Atayde:2021pgb,Esposito:2021ect,Koussour:2022irr,Koussour:2022jss,Koussour:2022wbi,Dixit:2022vyz,Sarmah:2023oum,Pradhan:2022dml,Bhar:2023xku, Bhar:2023yrf,Dimakis:2022rkd,Dimakis:2022wkj,Capozziello:2022tvv}, black holes \cite{DAmbrosio:2021zpm,Javed:2023qve,Nashed:2001im,Javed:2023vmb,Junior:2023qaq,Nashed:2001cp,Gogoi:2023kjt,Bahamonde:2022esv}, and wormholes \cite{Banerjee:2021mqk,Kiroriwal:2023nul,Mustafa:2023kqt,Godani:2023nep,Mishra:2023bfe,Hassan:2022ibc,Hassan:2022hcb,Parsaei:2022wnu,Sokoliuk:2022efj,Jan:2023djj}. Moreover, \cite{Wang:2021zaz,Lin:2021uqa,Calza:2022mwt} study particular $f(\mathbb{Q})$ solutions of spherically symmetric spacetime, whereas Heisenberg offers a thorough analysis of the theory in \cite{Heisenberg:2023lru}. { Moreover, an investigate of gravitational waves in $f(Q)$ gravity has been discussed in \cite{Capozziello:2024vix}. Analysis of the gravitational waves in different extensions of symmetric teleparallel, focusing on their speed and polarization has been carried out in \cite{Soudi:2018dhv}. A discussion of the  propagation velocity of the gravitational waves around Minkowski spacetime and their potential polarizations has been investigated in \cite{Hohmann:2018wxu}.}

In GR, researchers often present solutions with spherically symmetry  along with cylindrically symmetric, static metrics. These approaches offer various ways to achieve useful outcomes. The earliest cylindrically symmetric solutions in GR are Levi-Civita (LC) solutions \cite{Levi-Civita:1917pgo}. LC solutions, discovered shortly after Schwarzschild's solution, involve two independent parameters. Another significant solution within cylindrically symmetric spacetime in GR is described by cosmic strings, which are also known as topological defects \cite{Hiscock:1985uc,Linet:1986sr,Tian:1986zz,Zofka:2007bi}.  These solutions represent external effects caused by a cylindrically symmetric source.  Several modified theories, such as Brans-Dicke \cite{Delice:2006uz}, have been explored for this spacetime. These include $f(G)$, $f(R)$-gravity \cite{Azadi:2008qu}, $f(R,G)$, gravity \cite{Houndjo:2013us}, mimetic gravity \cite{Momeni:2015aea,Nashed:2018qag}, $f(R,\phi,X)$, gravity \cite{Zia:2019ddd} $f(T)$-gravity \cite{Houndjo:2012sz}, and the Einstein-Aether theory \cite{Chan:2021ivp,Malik:2022cyz}.  However, $f(\mathbb{Q})$-gravity has not been explored for this spacetime yet. This motivates us to study the static metric with cylindrical symmetry within coincident $f(\mathbb{Q})$ theory\footnote{The scope of this study is to derive a general static solution in the frame of $f(\mathbb{Q})$  and derive the corresponding form of the analytic function $f(\mathbb{Q})$. The  charged static case is out the scope of this study and will be  investigated elsewhere.}.

 The arrangements of this work are as follow: In Section \ref{sec2}, we review $f(\mathbb{Q})$ gravity and STEGR to set the scene for our computations. Section \ref{S3} addresses the four-dimensional field equations of $f(\mathcal{Q})$ using a metric potential with a flat horizon. We find a static solution in four dimensions that behaves similarly to dS/AdS at large distances. The key feature distinguishing this black hole from non-metricity solutions is the presence of a constant that causes the  $f(\mathcal{Q})$ to be different from zero or constant.  If we make this constant zero, then $f(\mathcal{Q})$ takes a constant value, and we recover   non-metricity theory. We investigate the singularities of this black hole in Section \ref{S4}, demonstrating that, compared to the non-metricity case, the singularity of the Riemann tensor squared  is much mild. Then, we generate a rotating solution with non-trivial $f(\mathcal{Q})$ and the non-metricity scalar $\mathbb{Q}$ using a coordinate transformation in Section \ref{S5}. Then, in Section \ref{S6}, we discuss the black hole thermodynamics,  and demonstrate its thermal stability, as indicated in Sections \ref{S3}. Finally, in the concluding section, we provide some final thoughts before concluding this study.

\section{$f(\mathbb{Q})$ Gravity}
\label{sec2}
\par
A mathematical concept called the connection establishes the covariant derivative and provides direction for tensor properties on the manifold.  In GR, the connection, determined by a symmetric connection called the Christoffel symbol or Levi-Civita connection, is based on Riemannian geometry.  On the other hand, relaxing the metricity condition leads to a more general connection with another component when its antisymmetric part is combined.

The specific form of this connection, known as an affine connection, is:
\begin{equation}
{\mathit \Gamma_{\phantom{\beta}\mu\nu}^{\beta}}{\mathcal =\left\{ _{\phantom{\beta}\mu\nu}^{\beta}\right\} +K_{\phantom{\beta}\mu\nu}^{\beta}+L_{\phantom{\beta}\mu\nu}^{\beta},}
\label{conexion}
\end{equation}
with  $\left\{ _{\phantom{\beta}\mu\nu}^{\beta}\right\}$ being the Christoffel symbol, which is consistent with the metric,
\begin{equation}
\mathit{ \left\{ _{\phantom{\beta}\mu\nu}^{\beta}\right\} =\frac{1}{2}g^{\beta\alpha}\left(\partial_{\mu}g_{\nu\alpha}+\partial_{\nu}g_{\alpha\mu}-\partial_{\alpha}g_{\mu\nu}\right)}.\label{christoffel}
\end{equation}
The second term, ${\mathit K_{\phantom{\beta}\mu\nu}^{\beta}}$, is called the contorsion, which is the antisymmetric part of the connection. It's determined by the torsion tensor. \begin{align}
\mathit{  T_{\phantom{\beta}\mu\nu}^{\beta}=2\varGamma_{\phantom{\beta}\left[\mu\nu\right]}^{\beta}=-T_{\phantom{\beta}\nu\mu}^{\beta}},\end{align}
\begin{equation}
\mathit { K_{\phantom{\alpha}\mu\nu}^{\beta}=\frac{1}{2}T_{\phantom{\alpha}\mu\nu}^{\beta}+T_{(\mu\phantom{\beta}\nu)}^{\phantom{\mu}\beta}}.\label{tns_tor} 
\end{equation}
The disformation tensor, ${\mathit L_{\phantom{\beta}\mu\nu}^{\beta}}$, is the last item in Eq.(\ref{conexion}).
\begin{equation}
\mathit{ L_{\phantom{\alpha}\mu\nu}^{\beta}=\frac{1}{2}\mathbb{Q}_{\phantom{\alpha}\mu\nu}^{\beta}-\mathbb{Q}_{(\mu\phantom{\beta}\nu)}^{\phantom{\mu}\beta}=L_{\phantom{\beta}\nu\mu}^{\beta}},\label{disf}
\end{equation}
which has the non-metricity tensor as its definition has the form:
\begin{equation}
\mathit{ \mathbb{Q}_{\beta\mu\nu}\equiv \nabla_{\beta}g_{\mu\nu}}.   \label{tns_nmetric}
\end{equation}
Now, it's clear that the affine connection \eqref{conexion} and the covariant derivative $\nabla_\mu$ are related.

To simplify the motion equations, it is appropriate to define the superpotential, whose  form is given by:
\begin{equation}
\mathit{P_{\phantom{\alpha}\mu\nu}^{\beta}=-\frac{1}{2}L_{\phantom{\alpha}\mu\nu}^{\beta}-\frac{1}{4}\left[\left(
\tilde{\mathbb{Q}}^{\beta}g_{\mu\nu}-\mathbb{Q}^{\beta}\right)g_{\mu\nu}+\delta_{(\mu}^{\beta}\mathbb{Q}_{\nu)}\right]},\label{superpot}
\end{equation}
with
\begin{align}
\mathit {\mathbb{Q}_{\alpha}=g^{\mu\nu}\mathbb{Q}_{\alpha\mu\nu}=\mathbb{Q}_{\alpha\phantom{\nu }\nu}^{\phantom{\alpha}\nu}, \qquad
\tilde{\mathbb{Q}}_{\alpha}=g^{\mu\nu}\mathbb{Q}_{\mu\alpha\nu}=\mathbb{Q}_{\phantom{\nu}\alpha\nu}^{\nu}},\end{align} being the non-metricity tensor's traces.
\par
So, by combining the non-metricity tensor \eqref{tns_nmetric} with the superpotential \eqref{superpot} through contraction, we provide a clearer expression for the non-metricity scalar.
\begin{equation}
\mathit{    \mathbb{Q}=-\mathbb{Q}_{\beta\mu\nu}P_{\phantom{\alpha}}^{\beta\mu\nu}}.\label{scalarQ}
\end{equation}
\par
In GR, the curvature tensor is determined by the relationship between Levi-Civita connection.
\begin{equation}
\mathit{R_{\phantom{\beta}\mu\alpha\nu}^{\beta}=\partial_{\alpha}\Gamma_{\phantom{\beta}\nu\mu}^{\beta}-
\partial_{\nu}\Gamma_{\phantom{\beta}\alpha\mu}^{\beta}+\Gamma_{\phantom{\beta}\alpha\rho}^{\beta}
\Gamma_{\phantom{\rho}\nu\mu}^{\rho}-\Gamma_{\phantom{\beta}\nu\rho}^{\beta}\Gamma_{\phantom{\rho}\alpha\mu}^{\rho}}.    \label{tns_Riem}
\end{equation}
The Ricci tensor can be obtained by performing a reduction within this tensor.
\begin{equation}
  \mathit{  R_{\mu\nu}=R_{\phantom{\beta}\mu\beta\nu}^{\beta}}.
\end{equation}
The Ricci scalar is found by summing up the components of the Ricci tensor.
\begin{equation}
 \mathit{ R= g^{\mu\nu}R_{\mu\nu}}.
\end{equation}

We can express the Riemann tensor \eqref{tns_Riem} differently by breaking it down using the affine connection.
\begin{equation}
\mathit{R_{\phantom{\beta}\alpha\mu\nu}^{\beta}=\overset{C}{R}{}_{\phantom{\beta}\alpha\mu\nu}^{\beta}+
\overset{C}\nabla_{\mu}V_{\phantom{\beta}\nu\alpha}^{\beta}-\overset{C}\nabla_{\nu}V_{\phantom{\beta}\mu\alpha}^{\beta}
+V_{\phantom{\beta}\mu\rho}^{\beta}V_{\phantom{\rho}\nu\alpha}^{\rho}-V_{\phantom{\beta}\nu\rho}^{\beta}V_{\phantom{\rho}
\mu\alpha}^{\rho}},\label{Tns_Riem_Trans}
\end{equation}
In this expression, we use the affine connection $\overset{C}{R}{}_{\phantom{\beta}\alpha\mu\nu}^{\beta}$ and the derivative $\overset{C}\nabla$ to describe $R_{\phantom{\beta}\alpha\mu\nu}^{\beta}$. The terms $ \overset{C}{\Gamma}{}_{\phantom{\beta}\alpha\mu}^\beta$ and \eqref{christoffel} involve quantities related to the Christoffel symbol.  $V^\beta_{\phantom{\beta}\mu\nu}$ is a tensor given by,
\begin{equation}
  {\mathit  V^\beta_{\phantom{\beta}\mu\nu}=K^\beta_{\phantom{\beta}\mu\nu}+L^\beta_{\phantom{\beta}\mu\nu}}.
\end{equation}
\par
Additionally, given a torsion-free connection $T_{\phantom{\alpha}\mu\nu}^{\beta}=0$, the relation \eqref{Tns_Riem_Trans} reduces to, given the proper contractions applied to the Riemann tensor,
\begin{equation}
\mathit{R=\overset{C}{R}-\mathbb{{Q}}+\overset{C}\nabla_{\beta}\left(\mathbb{Q}^{\beta}-\tilde{\mathbb{Q}}^{\beta}\right)},\label{scalar_Ric}
\end{equation}
where the Ricci scalar, expressed in terms of the Christoffel symbol, is $\overset{C}{R}$.
\par
We will thus find a more general method linking the Ricci scalar with the non-metricity scalar, starting from the teleparallel condition (TC) given by $R=0$, i.e. we will have a flat space-time that establishes the teleparallel geometries,
\begin{equation}
\mathit{\overset{C}{R}=\mathbb{Q}-\overset{C}\nabla_{\beta}\left(\mathbb{Q}_{\phantom{\beta}}^{\beta}-
\tilde{\mathbb{Q}}{}_{\phantom{\beta}}^{\beta}\right)}.
\label{scalar_Ric2}
\end{equation}
So, this relationship shows that a total derivative term or a boundary term separates the non-metricity scalar from the Ricci scalar
\begin{equation}
\mathit{B_\mathbb{Q}=\overset{C}\nabla_{\beta}\left(\mathbb{Q}^{\beta}-\tilde{\mathbb{Q}}^{\beta}\right)}.\label{boundary}
\end{equation}
Next, the following action presents a gravitational theory called STEGR. In this theory, the non-metricity tensor describes the gravitational interaction
\begin{equation}
\mathit{S_{\text{STEGR}}=\int\sqrt{-g}d^{4}x\Big[\mathbb{Q}+2\kappa^2\mathcal{L}_m\Big]}\label{action_Q}.
\end{equation}
Here $\mathcal{L}_m$ represents the Lagrangian of the matter field, and $\kappa^2=8\pi G/c^4$ stands for the gravitational constant\footnote{ {Charge is fundamentally considered a property of particles that relates to their interactions through fundamental forces, particularly electromagnetism. In classical physics, electric charge is treated as a conserved quantity in isolated systems, typically arising from symmetry principles in the underlying physical laws \cite{Landau:1975pou}.  Noether's theorem is a fundamental result in theoretical physics linking symmetries and conservation laws. It states that every differentiable symmetry of the action of a physical system corresponds to a conservation law. In the case of electric charge, gauge symmetry (specifically U(1) symmetry in quantum field theory) leads to the conservation of charge.  If there is an unknown or hidden symmetry in a physical system, Noether's theorem implies that it might give rise to a new conserved quantity, possibly linked to phenomena not yet fully understood \cite{Peskin:1995ev}. The charge itself can sometimes be a manifestation of a deeper, hidden symmetry not evident at lower energies or scales of the system. This is a typical line of reasoning in advanced physics models like Grand Unified Theories (GUTs) or theories beyond the Standard Model.
}}. It's important to highlight that according to relation \eqref{scalar_Ric2}, there exists a boundary term ($B_\mathbb{Q}$) that distinguishes the Einstein-Hilbert action of GR from the action of STEGR theory. This implies that STEGR provides an alternative description of GR.
\par
The STEGR theory can be expanded in a nonlinear manner by employing the following action:
\begin{equation}
\mathit{S_{\rm f_\mathbb{Q}}=\int\sqrt{-g}d^{4}x\Big[f\left(\mathbb{{Q}}\right)+2\kappa^2\mathcal{L}_{m}\Big]}\label{action_f(Q)}.
\end{equation}
where the non-metricity scalar $\mathbb{Q}$ can take on any function, represented by $f(\mathbb{Q})$. The  field equations of $f(\mathbb{Q})$ theory are obtained by varying the action \eqref{action_f(Q)} in relation to the metric \cite{BeltranJimenez:2019tme}.
\begin{equation}
\mathit{\frac{2}{\sqrt{-g}}\nabla_{\alpha}\left(\sqrt{-g}f_{\mathbb{Q}}\left(\mathbb{Q}\right)
P_{\phantom{\alpha}\mu\nu}^{\alpha}\right)+\frac{1}{2}g_{\mu\nu}f(\mathbb{Q})+f_{\mathbb{Q}}\left(\mathbb{Q}\right)
\left(P_{{\mu{\alpha}{{\nu}}}}\mathbb{Q}_{\nu}^{\phantom{\nu}\alpha\beta}-2\mathbb{Q}_{\alpha\beta\mu}
P_{\phantom{\alpha\beta}\nu}^{\alpha\beta}\right)=\kappa^2 \Theta_{\mu\nu}}.\label{eq_fie_f(Q)}
\end{equation}
Here, $\Theta_{\mu\nu}$ stands for the momentum-energy tensor. To make the equations simpler, we use the notation: $ f_{\mathbb{Q}}\equiv \frac{\partial f(\mathbb{Q})}{\partial\mathbb{Q}}$.
\par
To derive the linear equations of the STEGR theory, we perform the functional variation of equation \eqref{action_Q}, which is accomplished by differentiating with respect to \eqref{eq_fie_f(Q)}. This process involves $f(\mathbb{Q})$ and $\mathbb{Q}$.
\par
According to \cite{Lin:2021uqa,Zhao:2021zab,DAmbrosio:2021zpm}, we can rewrite the equations of motion \eqref{eq_fie_f(Q)} in a more convenient manner as:
\begin{equation}
\zeta_{\mu\nu}\equiv \mathit{f_{\mathbb{Q}}\left(\mathbb{Q}\right)G_{\mu\nu}-\frac{1}{2}g_{\mu\nu}\left[f(\mathbb{Q})-f_{\mathbb{Q}}
\left(\mathbb{Q}\right)\mathbb{Q}\right]+2f_{\mathbb{Q} \mathbb{Q}}\left(\mathbb{Q}\right)P_{\phantom{\alpha}\mu\nu}^{\alpha}\partial_{\alpha}\mathbb{Q}=
\kappa^2\Theta_{\mu\nu}.}\label{eq_f(Q)}
\end{equation}
Here, $f_{\mathbb{Q} \mathbb{Q}}$ denotes the second derivative of the function $f(\mathbb{Q})$ with respect to $\mathbb{Q}$, which represents the Einstein tensor as described by \eqref{christoffel}.
\par
We make the assumption that the matter content is disappearing in this study, i.e., $\Theta_{\mu\nu}=0$ .
\section{Exact  black holes in f({\cal Q})  theory}\label{S3}
We can make a 4-dimensional spacetime cylinder by using the equations from the $f(\mathbb{Q})$ theory, Eq. (\ref{eq_f(Q)}). When we do this, the metric in cylindrical coordinates becomes ($t$, $r$, $\xi_1$, $\xi_2$), as explained in \cite{Capozziello:2012zj}:
\begin{eqnarray}\label{met1} ds{}^2=\mu(r)dt^2-\frac{1}{\nu(r)} dr^2-r^2(d\xi^2_1+d\xi^2_2).\end{eqnarray}
In this setup, we represent the unknown functions $\nu(r)$ and $\mu(r)$, which depend on the radial coordinate $r$. { To find the non-metricity scalar}, we use equations (\ref{met1}) and (\ref{scalarQ})\footnote{To keep things simple, we'll use the following shorthand: $\mu(r)$ is represented as $\mu$, $\nu(r)$ is represented as $\nu$. The derivative of $\mu$ or $\nu$  with respect to $r$ are denoted as $\mu'$ and $\nu'$.}
\begin{equation}\label{df}
{\mathbb{Q}(r)}=-\frac{2\nu[\mu+\mu']}{r^2 \mu}.
\end{equation}
The equations of motion (\ref{eq_f(Q)}) can be obtained by applying Eq. (\ref{met1}) when ${{\cal T}^{{}^{{}^{^{}{\!\!\!\!\scriptstyle{em}}}}}}^\nu_\mu=0$, the following non-zero components are obtained:

\begin{align}\label{df1}
&\zeta_t{}^t\equiv\frac{1}{2r^4\mu^2}\left[8 \nu {}^2r^2f_{\mathbb{Q}\mathbb{Q}}[\mu'^2-\mu \mu '']+2r\mu \nu \mu '[4f_{\mathbb{Q}\mathbb{Q}}\{\nu -r\nu'\}+r^2f_\mathbb{Q}]+\mu \left\{2r\mu \nu'\left[r^2f_\mathbb{Q}-4f_{\mathbb{Q}\mathbb{Q}}\nu \right]+\mu \left[r^4\{f(\mathbb{Q})-2\Lambda\}
\right.\right.\right.\nonumber\\
& \left.\left.\left.+4r^2\nu f_\mathbb{Q}+4
4\nu {}^2f_{\mathbb{Q}\mathbb{Q}}\right]\right\}\right]=0\,,\nonumber\\
&\zeta_r{}^r\equiv\frac{\{f(\mathbb{Q})-2\Lambda\} r^2 \mu +4rf_\mathbb{Q}\nu \mu'+4f_\mathbb{Q}\mu\nu }{2r^2h}=0\,,\nonumber\\
&\zeta_{\xi_1}{}^{\xi_1}=\zeta_{\xi_2}{}^{\xi_2}\equiv\frac{1}{4r^4\mu ^3}\left\{2r^2\nu \mu \mu ''[\mu (r^2f_\mathbb{Q}-4f_{\mathbb{Q}\mathbb{Q}}\nu )-2r\nu \mu'f_{\mathbb{Q}\mathbb{Q}}]+4f_{\mathbb{Q}\mathbb{Q}}r^3\nu {}^2\mu '^3-r^2\mu \nu \mu'^2[r^2f_\mathbb{Q}+4f_{\mathbb{Q}\mathbb{Q}}(r\nu'-3\nu )]+r\mu ^2\mu '
\right.\nonumber\\
&\left.
\times\left[r\mu '_1(r^2f_\mathbb{Q}-12f_{\mathbb{Q}\mathbb{Q}}\nu )+5r^2f_\mathbb{Q}\nu +16f_{\mathbb{Q}\mathbb{Q}}\nu {}^2\right]+2\mu ^2\left[r\mu \nu'
(r^2f_\mathbb{Q}-4\nu f_{\mathbb{Q}\mathbb{Q}})+\mu \left(r^4\{f(\mathbb{Q})-2\Lambda\}+2r^2\nu f_\mathbb{Q}+8f_{\mathbb{Q}\mathbb{Q}}\nu {}^2\right)\right]\right\}=0\,.
\end{align}
To solve the differential equation system mentioned earlier, we utilize the chain rule along with the subsequent data:
\begin{eqnarray}\label{dfg}
&&f({\mathbb Q})=f(r),\nonumber\\
& & f_{ \mathbb  Q}=\frac{df({\mathbb   Q})}{d{\mathbb  Q}}=\frac{df(r)}{dr}\frac{d r}{d{\mathbb  Q}}=\frac{r^3 \mu^2 f'}{2[r^2\mu \mu'\nu'+r\mu^2\nu'+r^2\mu \nu \mu''-r\mu \nu \mu'-2\mu^2\nu-r^2\nu \mu'^2]},\nonumber\\
& & f_{{\mathbb  Q}{\mathbb  Q}}=\frac{df_{\mathbb Q}}{d{\mathbb Q}}={ \frac{d}{dr}\Bigg(\frac{df(r)}{dr}\frac{d r}{d{\mathbb Q}}\Bigg)\frac{d r}{d{\mathbb  Q}}}=\frac{r^5 \mu^3}{4[r^2\nu[\mu \mu''-\mu'^2]+r\mu \mu'[r\nu'-\nu]+\mu^2(r\nu'-2\nu)]^{^{^3}}}\times \nonumber\\
& &\times\Big[ r \mu f''\Big(r^2\mu \nu \mu''-r^2\nu \mu'^2+r\mu \mu'\{r\nu'-\nu\}+\mu^2\{r\nu'-2\nu\}\Big)-r^3f'\mu^2 \nu \mu'''+f'\Big(3r^3\mu \nu \nu'\nonumber\\
& &-2r^2\mu^2\mu''[r\nu'-\nu]-r^2\mu^2\mu''_1[r\mu'+\mu]-2r^3\nu \mu'^3
+2r^2\mu \mu'^2[r\nu'-\nu]+2r\mu^2\mu'[r\nu'-\nu]\nonumber\\
& &+2\mu^3[2r\nu'-3\nu]\Big)\Big] ,\nonumber\\
& &
\end{eqnarray}
where $f'=\frac{df(r)}{dr}$ and  $f''=\frac{d^2f(r)}{dr^2}$.
A general $d$-dimension  solution  of Eq. (\ref{df1}), after using Eq. (\ref{dfg}), takes the form
\begin{eqnarray} \label{sol}
 &&\mu(r)=r^2c_1 +\frac{c_2}{r}, \quad  \nu(r)=\left[1-\frac{c_3e^{\frac{c_3}{r}}}{r}\right]\left[r^2c_1 +\frac{c_2}{r}\right]\equiv\mu(r)\left[1-\frac{c_3e^{\frac{c_3}{r}}}{r}\right], \nonumber\\
  &&f(r)=c_4\sqrt{1-\frac{c_3e^{\frac{c_3}{r}}}{r}}+2\Lambda,
\end{eqnarray}
where $c_i,\, \, i=1\cdots 4$  are  integration constants.  It's important to note that when $c_3$ and $c_4$ are both zero, we achieve
\begin{eqnarray} \label{sol1}
 &&\mu(r)=\nu(r)=r^2c_1 +\frac{c_2}{r}, \qquad  f(r)=2\Lambda,
\end{eqnarray}
which corresponds to the linear case of the $f(\mathbb{Q})$ scenario \cite{Nashed:2006yw,Nashed:2005kn,Nashed:2003ee}.
\section{Physical properties  of the  solution}\label{S4}
This section examines the key physics of the previously found solution.\vspace{0.2cm}\\
\underline{Non-metricity  scalar:}\vspace{0.2cm}\\
By plugging Eq.~(\ref{sol}) into Eq.~(\ref{df}), we find the non-metricity scalar as follows:
\begin{eqnarray} \label{Tor1} &&{\mathbb Q}=6c_1\left(1-\frac{c_3e^\frac{c_3}{r}}{r}\right)\Longrightarrow\nonumber\\
&& r= \frac{c_3}{
 {LambertW} \left[ \frac {
 \left( {\mathbb Q}+6\,c_1 \right)}{6{c_1}} \right]}.\end{eqnarray}
According to the first equation of Eq. (\ref{Tor1}), when $c_3=0$, the non-metricity $Q$ can have a constant value, allowing us to return to the linear case of  $f(\mathbb{Q})$ theory.  Using Eq. (\ref{Tor1}) in the form of $f(r)$ given by Eq. (\ref{sol}) we get the form of $f(\mathbb{Q})$ related to solution (\ref{sol}) in the form:
\begin{align}\label{Q1}
f \left( {\mathbb Q}\right) =2\,{\Lambda}+\sqrt {-{\frac {Q}{
{6c_1}}}}{c_3}\,.
\end{align}
Equation (\ref{Q1}) shows that the constant $c_1$ which is related to the cosmological constant may take a negative value or $Q$ has a negative value so that $f(\mathbb{Q})$ becomes a real function.
\vspace{0.2cm}\\
\underline{Metric:}\vspace{0.2cm}\\
{ The metric given  by (\ref{met1}) takes the form:}
\begin{eqnarray}\label{met} ds{}^2=\mu dt^2-\frac{1}{\nu} dr^2-r^2(d\xi^2_1+d\xi^2_2).\end{eqnarray}
Using Eq. (\ref{sol})  in Eq. (\ref{met}) we get
\begin{eqnarray}\label{met11} ds{}^2=\Biggl(r^2\Lambda -\frac{M}{r}\Biggr)dt^2-\frac{dr^2}{(1-\frac{c_3e^{\frac{c_3}{r}}}{r})(r^2\Lambda  -\frac{M}{r})} -r^2(d\xi^2_1+d\xi^2_2),\end{eqnarray} where we have put $c_1=\Lambda$ and $c_2=-M$.
When $c_3$ equals zero, Eq. (\ref{met11}) reveals that we revert to the linear scenario of $f(\mathbb{Q})$. Furthermore, as shown in Eq. (\ref{Tor1}), the non-metricity ${\mathbb Q}$ becomes a constant value.   In spherically symmetric spacetime, the temporal ansatz is equal to the inverse of the radial one.  However, this doesn't fulfill Eq. (\ref{met11}) unless $c_3$ equals zero.  However, the two components $g_{tt}$ and $g_{rr}$ share equal event and killing horizons.\vspace{0.2cm}\\

\underline{Singularity:}\vspace{0.2cm}\\
In this subsection, we will focus on the case of four dimensions.  To study singularities, it's crucial to find the values of $r$ where the functions $\mu(r)$ and $\nu(r)$ reach zero or infinity.  Because singularities might correspond to coordinate ones, the usual approach involves analyzing various invariants constructed using the non-metricity and Levi-Civita affine connection.  We reproduce the following invariants from the curvature and non-metricity of the metric (\ref{met1}):
 \begin{eqnarray}
 && R^{\mu \nu \lambda \rho}R_{\mu \nu \lambda \rho} =\frac{3}{{r}^{10}}\left\{ c_3^{2} \left[ 5\,{r}^{8}{\Lambda}^{2}-2\,c_3{r}^{7}{\Lambda}^{2}+c_3^{2}{r}^{6}{\Lambda}^{2}+{r}^{5}\Lambda\,M
 -c_3^{2}\Lambda\,M\,{r}^{3}+{\frac {27\,M^{2}{r}^{2}}{4}}+\frac{7\,c_3rM^{2}}{2}+\frac{3\,M^{2}c_3^{2}}{4} \right]   e^{\frac{2 c_3}{r}}\right.\nonu
 &&\left.+4\, \left( c_3{\Lambda}^{2}{r}^{6}-3{r}^{7}{\Lambda}^{2}-\frac{{r}^{4}\Lambda\,M}{2}-\frac{\,c_3\Lambda\,M\,
 {r}^3}{2}-\frac{5\,r{M}^{2}}{2}-\frac{\,c_3{M}^{2}}{2} \right) c_3{r}^{2} e^{\frac{ c_3}{r}}+8\,{r}^2{\Lambda}^{2}+4\,{r}^{4}{M}^{2} \right\} \nonu
&&\approx 24\Lambda{}^2-\frac{36c_3\Lambda{}^2}{r}-\frac{9c_3{}^2\Lambda{}^2}{r^2}+{\cal O}\left(\frac{1}{r^3}\right) \qquad \mbox{as } r\rightarrow \infty,\nonu
   &&R^{\mu \nu}R_{\mu \nu} =\frac{3}{ {r}^{10}}\, \left\{ {c_3}^{2} \left( 7\,{r}^{8}{\Lambda}^{2}-4\,c_3{r}^{7}{\Lambda}^{2}+{c_3}^{2}{r}^{6}{\Lambda}^{2}+2\,{r}^{5}\Lambda\,M+
   c_3{r}^{4}\Lambda M-{c_3}^{2}\Lambda\,M\,{r }^{3}+\frac{3{M}^{2}}{8}[r^2+2c_3r+{c_3}^{2}] \right)  e^{{\frac{2c_3}{r}}}\right.\nonu
 &&\left.+6\, \left( {r}^{3}\Lambda\,c_3-3\,{r}^{4}\Lambda-\frac{M\,r}{n-2}+\frac{c_2\,c_3}{2} \right)c_3{r}^{5}\Lambda\,{e^{{ \frac {c_3}{r}}}}+12{r}^{10}{\Lambda}^{2} \right\}\approx 36\Lambda{}^2-\frac{54c_3\Lambda{}^2}{r}-\frac{15c_3{}^2\Lambda{}^2}{r^{2}}+{\cal O}\left(\frac{1}{r^3}\right) \qquad \mbox{as } r\rightarrow \infty,\nonu
  &&R =\frac{3}{r^5} \left\{  \left[ 3\,{r}^{4}\Lambda-{r}^3\Lambda\,c_3+\frac{\,M\,r}{2}+\frac{M\,c_3}{2} \right] c_3{{ e}^{{\frac {c_3}{r}}}}-4
\,{r}^5\Lambda \right\}
\approx -12\Lambda+\frac{9c_3\Lambda}{r}-\frac{6c_3{}^2\Lambda}{r^2}+{\cal O}\left(\frac{1}{r^3}\right) \qquad \mbox{as } r\rightarrow \infty,\nonu
&&{\cal Q}^{\mu \nu \lambda}{\cal Q}_{\mu \nu \lambda}=- \frac{1}{\left( {r}^{2}\Lambda-\frac{M}{r} \right){r}^{8} \left( 1-\frac{c_3}{r}e^{{\frac{ 2c_3}{r}}} \right)}\left\{ c_3^{2} \left( {M}^{2}c_3^{2}+13\,{r}^{8}{\Lambda}^{2}+13\,{M}^{2}{r}^{2}-2\,c_3{r}^{7}{\Lambda}^{2}+4\,c_3r{M}^{2}
+c_3^{2}{r}^{6}{\Lambda}^{2}
+2\,c_3^{2}\Lambda\,c_2\,{r}^{3}\right.\right.\nonu
 &&\left.\left.-2\,c_3{r}^{4}\Lambda\,M-8\,{r}^{5}\Lambda\,M \right)   e^{{\frac{ 2c_3}{r}}} +4\,c_3{r}^{2} \left( c_3{\Lambda}^{2}{r }^{6}-7\,{r}^{7}{\Lambda}^{2}+\frac{7{r}^{4}\Lambda\,M}2-\frac{c_3\Lambda\,M \,{r}^{3}}{2}-\frac{11r{M}^{2}}{2}-\frac{c_3{M}^{2}}{2} \right)e^{{\frac{ c_3}{r}}}\right.\nonu
 &&\left.-8\,{r}^{7}\Lambda\,M+16\,{r}^{10 }{\Lambda}^{2}-10\,{r}^{4}{M}^{2} \right\}  \approx 16\Lambda-\frac{12c_3\Lambda}{r}-\frac{7c_3{}^2\Lambda}{r^2}+{\cal O}\left(\frac{1}{r^3}\right) \qquad \mbox{as } r\rightarrow \infty,\nonu
&&{\cal P}^{\mu \nu \lambda}{\cal P}_{\mu \nu \lambda}=-\frac{1}{\left( {r}^{3}\Lambda-M\right){r}^{6} \left( r-c_3e^{{\frac {c_3}{r}}}\right)}\left\{ 3\,c_3^{2} \left( {M}^{2}c_3^{2}+33\,{r}^{8}{\Lambda}^{2}+29\,{M}^{2}{r}^{2}+2\,c_3{r}^{7}{\Lambda}^{2}
+6\,c_3r{M}^{2}+c_3^{2}{r}^{6}{\Lambda}^{2}-2\,c_3^{2}\Lambda\,M{r}^{3}\right.\right.\nonu
 &&\left.\left.
-8\,c_3{r}^{4}\Lambda\,M-38\,{r}^{ 5}\Lambda\,M \right) e^{{\frac {2c_3}{r}}}-12\,c_3{r}^{2} \left( 16\,{r}^{7}{\Lambda}^{2}-17\,{r}^ {4}\Lambda\,M-c_3\Lambda\,M\,{r}^{3}+13\,r{M}^{2}+c_3{M}^{2} \right)  e^{{\frac {c_3}{r}}}+96\,{r}^{7}\Lambda\,c_2+96\,{r}^{10}{\Lambda}^{2}\right.\nonu
 &&\left.+72\,{r}^{4}{M}^{2} \right\} \approx 3\Lambda-\frac{3c_3\Lambda}{r}+\frac{93c_3{}^2\Lambda}{32 r^2}+{\cal O}\left(\frac{1}{r^3}\right) \qquad \mbox{as } r\rightarrow \infty, \nonu
&&{\mathbb Q}=6\Lambda\left(1-\frac{c_3e^\frac{c_3}{r^{n-3}}}{r^{n-3}}\right)\approx -6\Lambda+\frac{6c_3\Lambda}{r}+\frac{6c_3{}^2\Lambda}{r^2}+{\cal O}\left(\frac{1}{r^3}\right) \qquad \mbox{as } r\rightarrow \infty,\end{eqnarray}

 The invariants suggest a singularity at $r=0$. Evaluating the invariants at $r=0$, we find ${\it K=Q^{\mu \nu \rho}Q_{\mu \nu \rho}=P^{\mu \nu \rho}P_{\mu \nu \rho}\sim r^{-1}}$. This is in contrast to the non-metricity black hole solutions of Einstein-Maxwell theory, i.e., the linear case of $f(\mathbb{Q})$ construction, where ${\it K=Q^{\mu \nu \rho}Q_{\mu \nu \rho}=P^{\mu \nu \rho}P_{\mu \nu \rho}\sim r^{-6}}$.   This suggests that the black hole's singularity is not as severe as it would be in the uncharged case in non-metricity. \vspace{0.1cm}\\


\section{Rotating solution in  $f({\cal Q})$ theory }\label{S5}

 The next step will be to solve the field equations (\ref{eq_f(Q)}) for a rotating black hole with two rotation parameters.  We limit ourselves for { this purpose using the static solution provided by Eq. (\ref{sol}).} Utilizing the $\phi_i$ rotation parameters and the ensuing transformations:
\begin{equation} \label{t1}
\bar{\xi}_{i} =-\Omega~ {\xi_{i}}+\frac{ \phi_i}{l^2}~t,\qquad \qquad \qquad
\bar{t}=
\Omega~ t-\sum\limits_{i=1}^{2}\phi_i~ \xi_i\,.
\end{equation}
Here,  $\Omega$ is defined  as:
\[\Omega:=\sqrt{1-\sum\limits_{j=1}^{{2}}\frac{\phi_j{}^2}{l^2}}.\]

The metric corresponds to   (\ref{t1}) is
\begin{eqnarray}
\label{m1}
    ds^2=\mu(r)\left[\Omega d{\bar {t}}  -\sum\limits_{i=1}^{2}  \phi_{i}d{\bar {\xi_1}}
\right]^2-\frac{dr^2}{\nu(r)}-\frac{r^2}{l^4}\sum\limits_{i=1 }^{2}\left[\phi_{i}d{\bar {t}}-\Omega l^2 d{\bar{\xi}}_i\right]^2-\frac{r^2}{l^2}\left(\phi_{1}d{\bar {\xi}}_2-\phi_{2}d{\bar {\xi}}_2\right)^2,
\end{eqnarray}
where $0 \leq \Omega_{i}< 2\pi$, $i=1,2$, and $-\infty < t < \infty$. Here, we observe that the static configuration (\ref{met}) can be recovered as a special case of the general metric previously mentioned, provided that the rotation parameters $\phi_j$ are chosen to be vanished.

\section{Thermodynamics of solution  (\ref{sol})}\label{S6}
 The definition of the Hawking temperature is \cite{PhysRevD.86.024013,Sheykhi:2010zz,Hendi:2010gq,PhysRevD.81.084040}:
  \begin{equation}
T_2 = \frac{\mu'(r_2)}{4\pi},
\end{equation}
when $\mu'(r_2)\neq 0$ is satisfied and the event horizon is at $r = r_2$.  For $f({\cal Q})$ gravitational theory, the Bekenstein-Hawking entropy is given by \cite{2013JCAP...11..060S,2013Ap&SS.344..259B}
\begin{equation}\label{ent}
S(r_2)=\frac{A}{4}\frac{df({\cal Q}_2)}{d{\cal Q}_2}=\frac{A}{4}\frac{df(r_2)}{dr_2}\frac{dr_2}{d{\cal Q}(r_2)},
\end{equation}
where the event horizon's area is denoted by $A$.
Finally, the following is the determination of the Gibbs free energy \cite{Zheng:2018fyn,Kim:2012cma}:
\begin{equation}
\label{gib}
G_2= E_2- T_2S_2\, ,
\end{equation}
where the temperature, entropy, and quasilocal energy at the event horizon are denoted, respectively, by $T(r_2)$, $S(r_2)$, and $E(r_2)$.

\begin{figure}[ht]
\centering
\subfigure[~The behavior of $g_{rr}$]{\label{fig:1a}\includegraphics[scale=0.35]{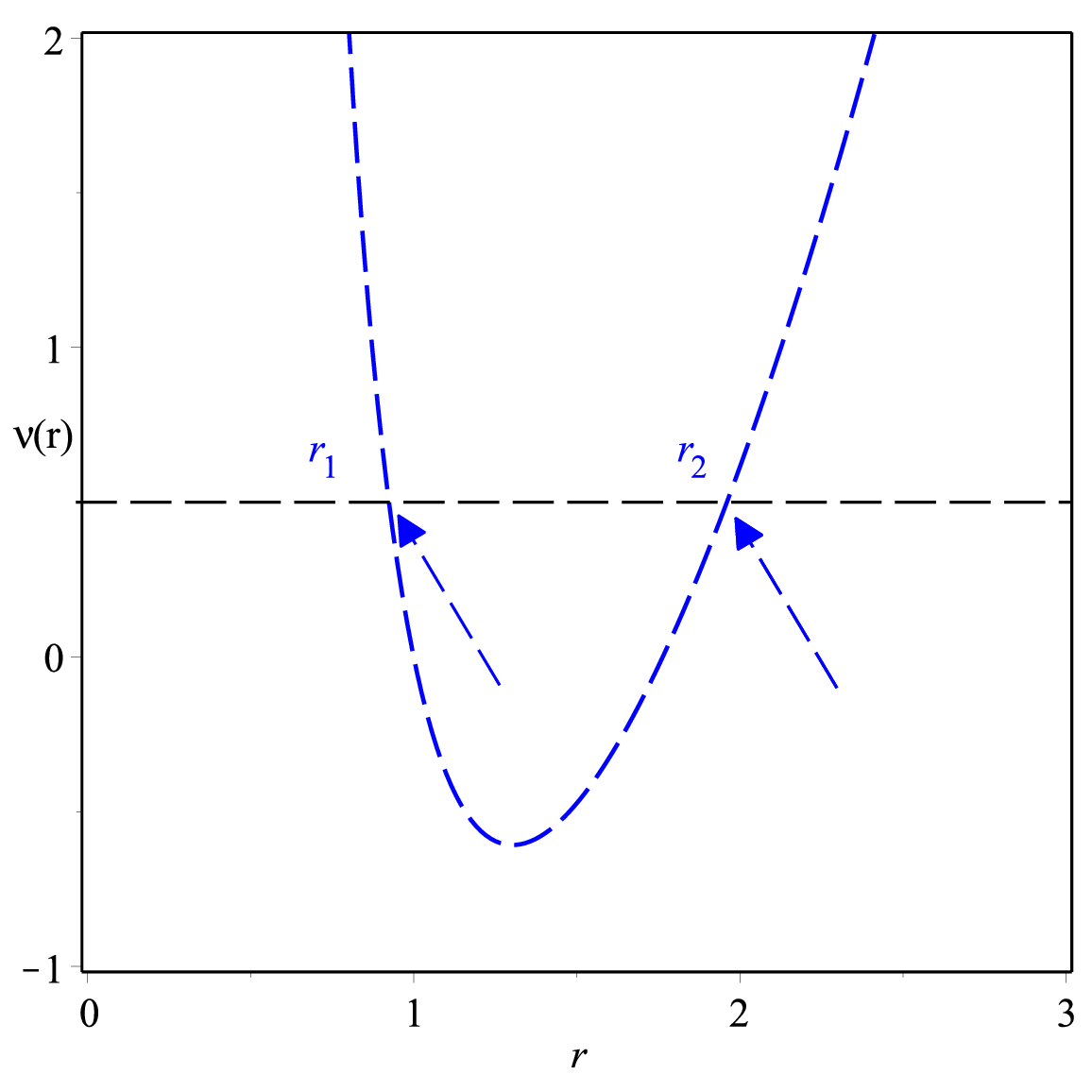}}\hspace{0.5cm}
\subfigure[~The behavior of  $Q$]{\label{fig:1b}\includegraphics[scale=0.35]{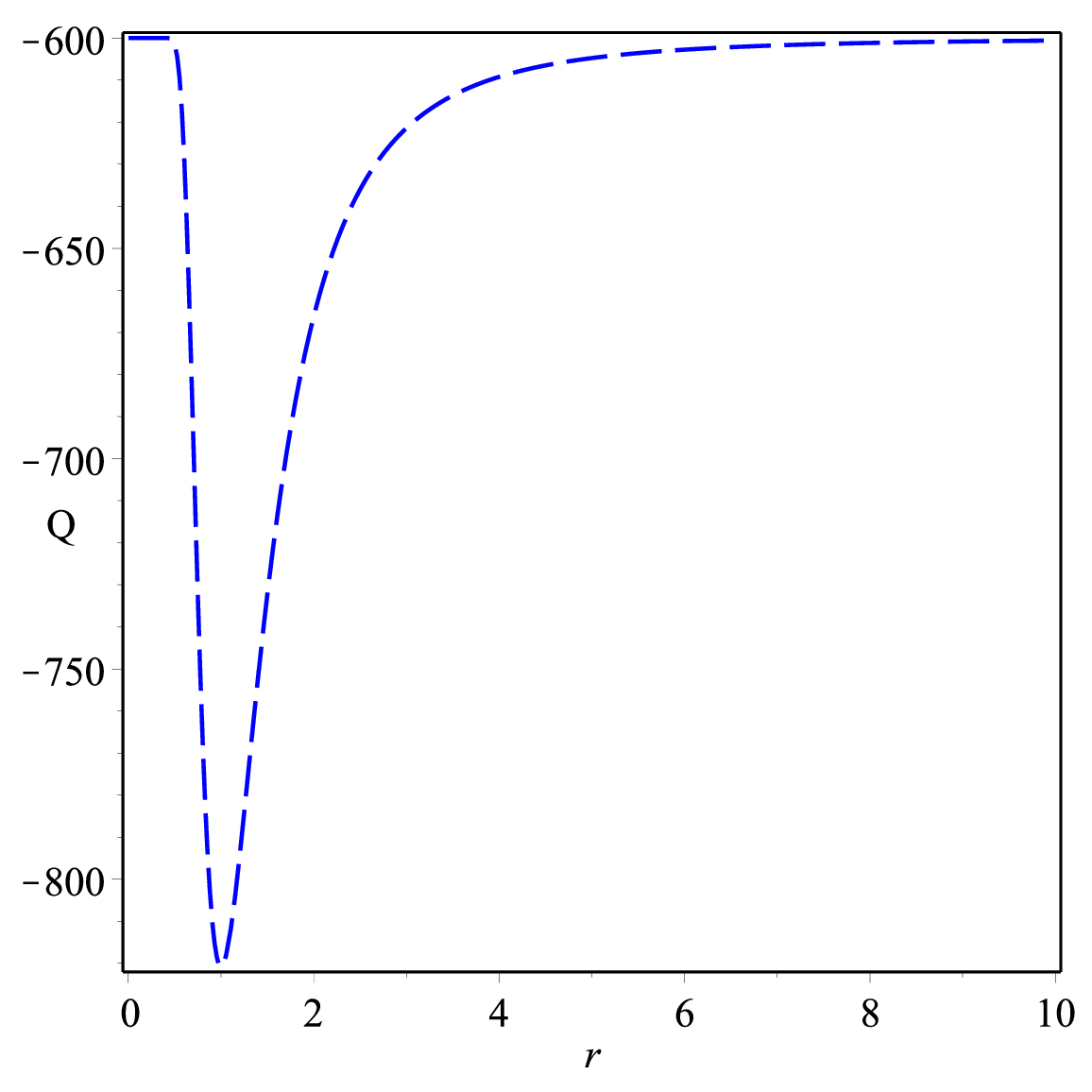}}\hspace{0.5cm}
\subfigure[~The behavior of the $f(r)$ and the radial coordinate $r$]{\label{fig:1c}\includegraphics[scale=0.35]{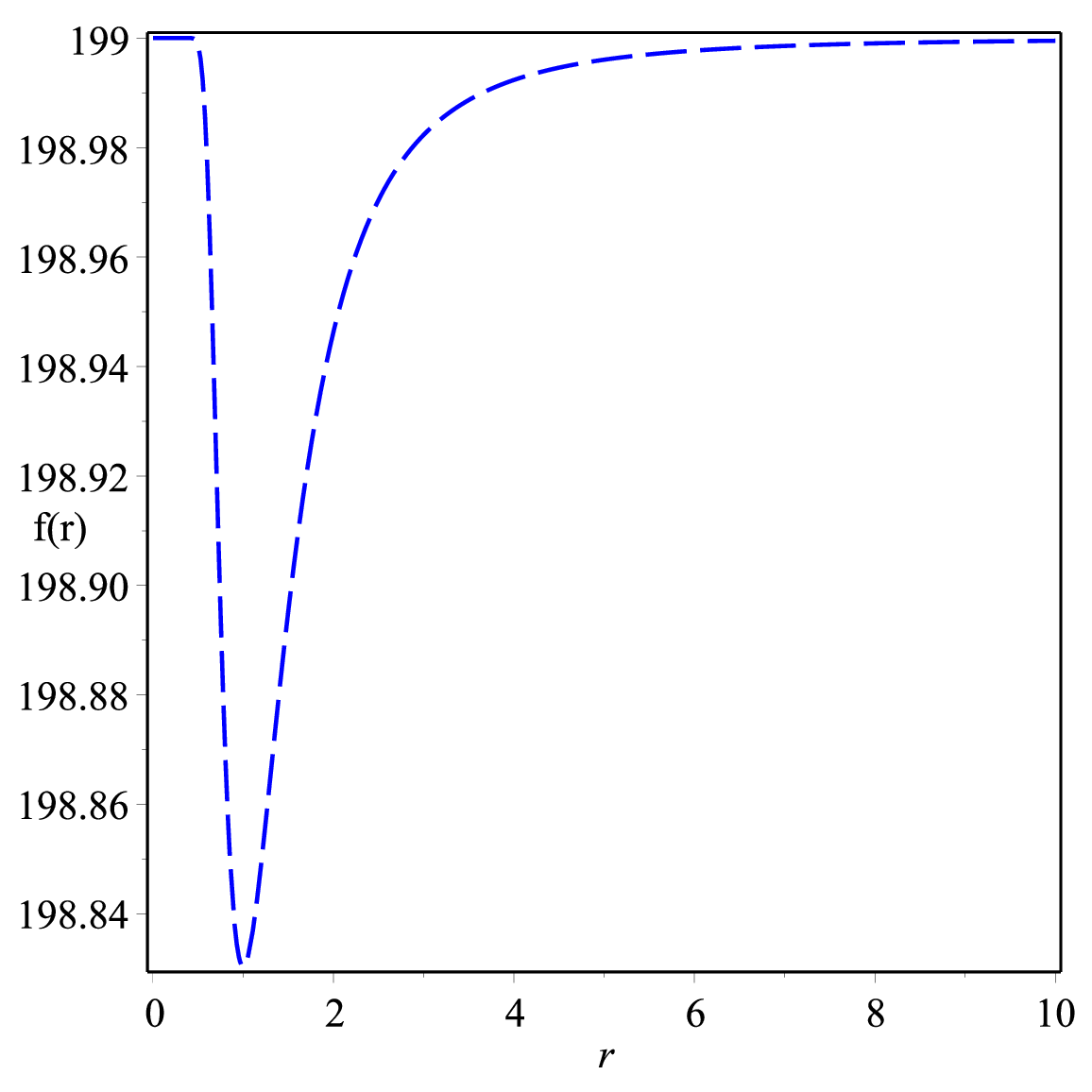}}\hspace{0.5cm}
\subfigure[~The behavior of the function $f(\mathbb{Q})$ and the non-metricity $Q$]{\label{fig:1d}\includegraphics[scale=0.35]{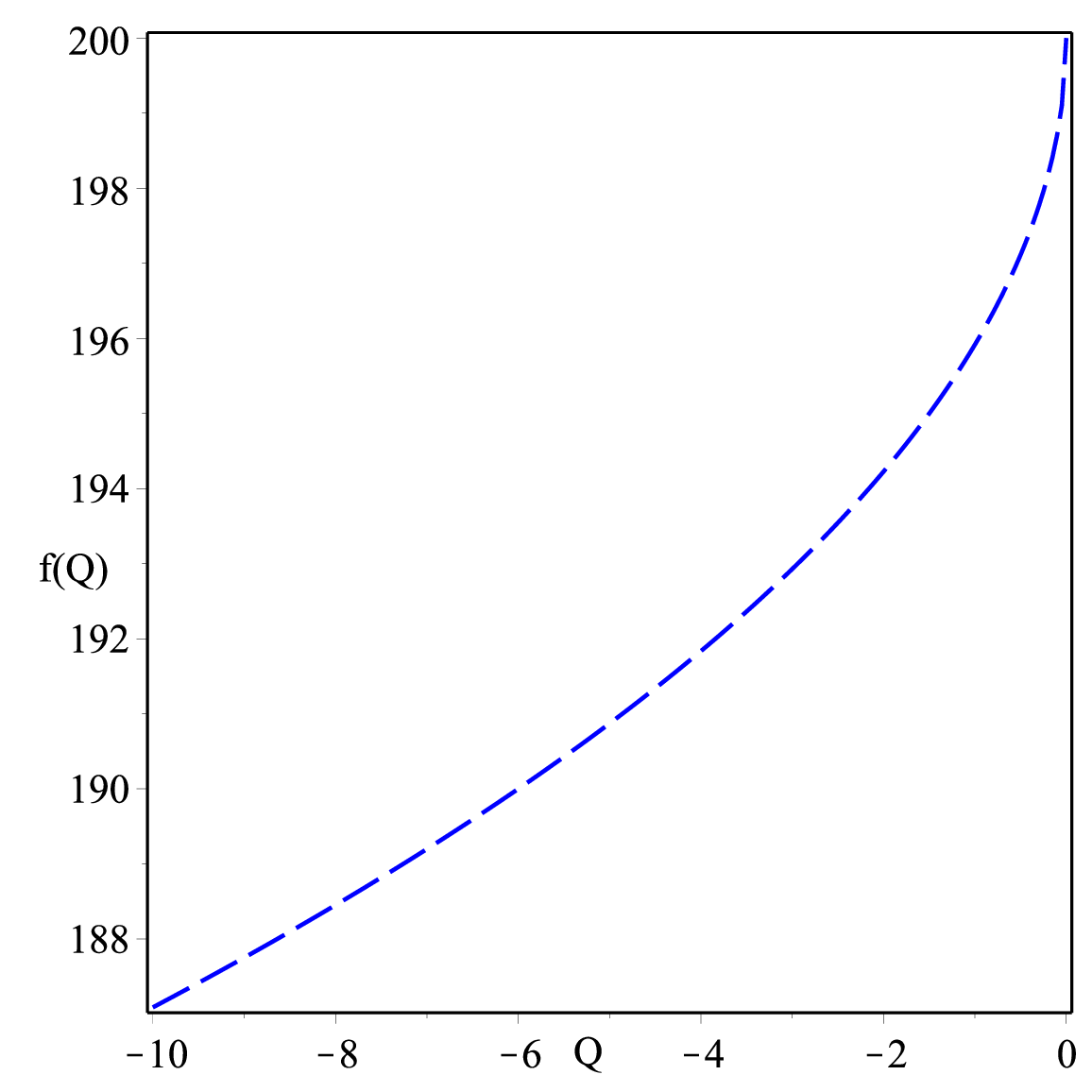}}
\caption{{\it{(a)The four-dimensional $f(\mathbb{Q})$ gravity metric function $\nu(r)$, where $r_2$ represents  the outer Cauchy horizon of a black hole. (b) The connection between the black hole's radial coordinate $r$ and the non-metricity $Q$ (\ref{met}).(c) The relationship between the radial coordinate $r$ and the black hole's $f(r)$ (\ref{met}).(d) The relationship between the non-metricity $Q$ and the black hole's $f(\mathbb{Q})$ (\ref{met}). The black hole's parameters (\ref{met}) all have the following numerical values in relativistic units: $\Lambda=100, c_4=-1,c_3=-100$ and $M=100$. }}}
\label{Fig:1}
\end{figure}

To explain the above thermodynamical quantities for the black hole solution (\ref{sol}), we begin with the constraint $\mu(r) = 0$. This gives
\begin{eqnarray} \label{m33}
&& {M}_{{}_{{}_{{}_{{}_{\tiny Eq. (\ref{sol})}}}}}=r{}^3\Lambda.
\end{eqnarray}
 Equation (\ref{m33}) states that the black hole's total mass is a function of the horizon radius. { Figure \ref{fig:1a} illustrates the relationship between the   $\nu(r)$ and  $r$, illustrating the potential horizons of the black hole.} Additionally, figure \ref{fig:1b} shows the relationship between the radial coordinate $r$ and the non-metricity $Q$ of the black hole (\ref{met}).  For the black hole (\ref{sol}), we illustrate the function $f(r)$ in figure \ref{fig:1c}, which exhibits positive behavior. In figure \ref{fig:1d}, we finally plot the relation between $f(\mathbb{Q})$ and $Q$, which likewise exhibits positive behavior.
\begin{figure}[ht]
\centering
\subfigure[~The entropy ]{\label{fig:2a}\includegraphics[scale=0.35]{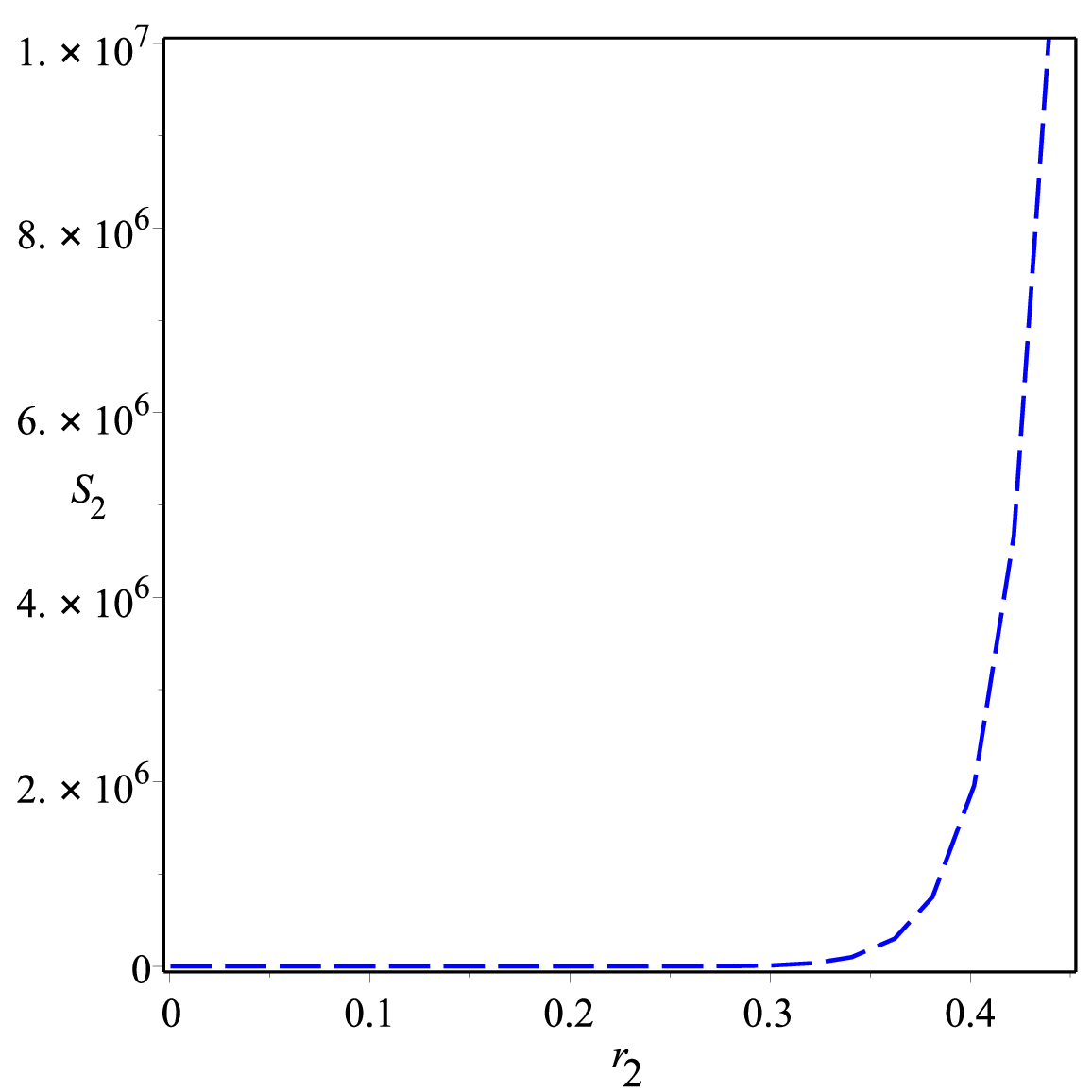}}\hspace{0.5cm}
\subfigure[~The Hawking temperature]{\label{fig:2b}\includegraphics[scale=0.35]{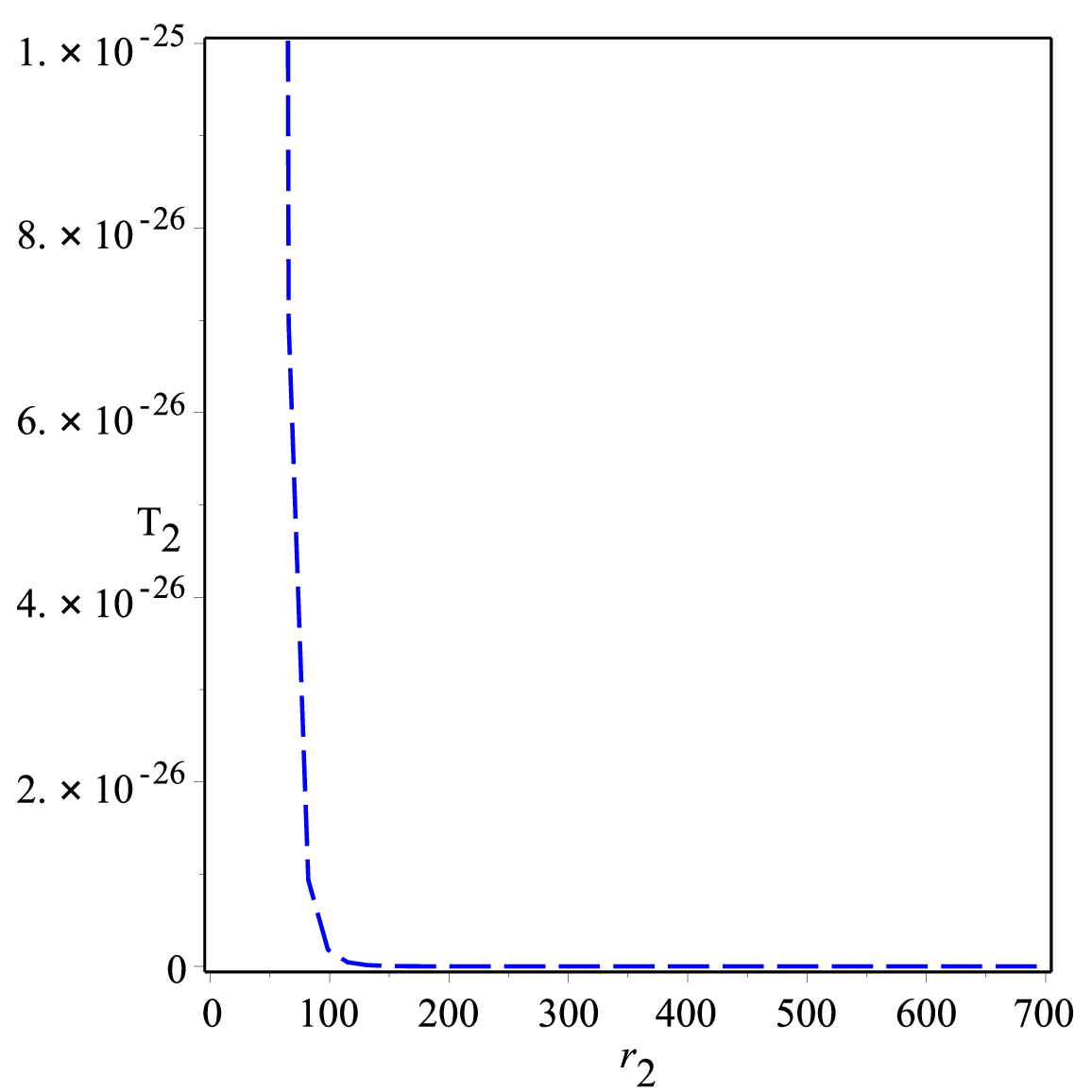}}\hspace{0.5cm}
\subfigure[~The Gibbs function]{\label{fig:2c}\includegraphics[scale=0.35]{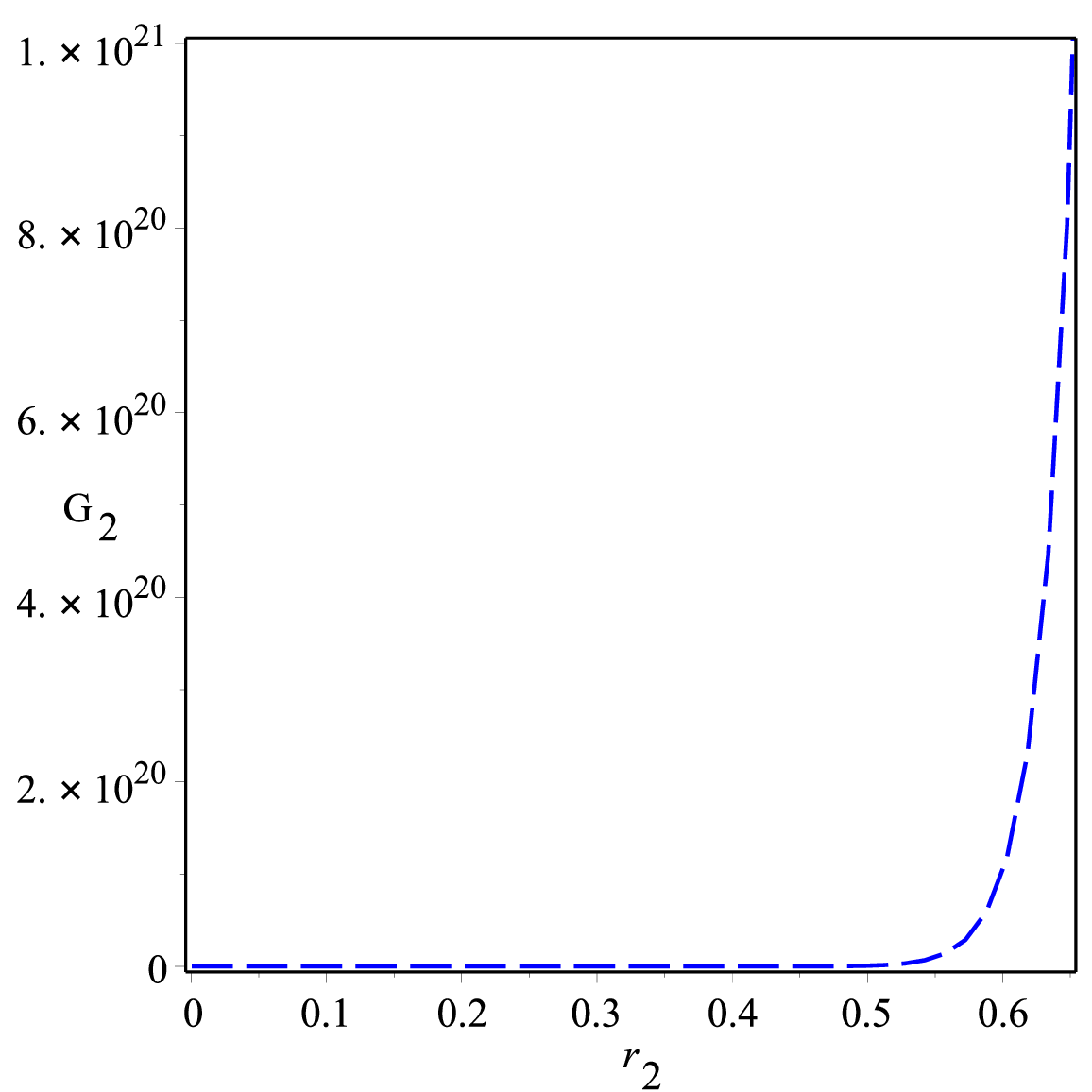}}
\caption{{\it{(a) The entropy of solution (\ref{sol}).(b) The Hawking temperature. (c) The Gibbs function. }}}
\label{Fig:3}
\end{figure}

Additionally, $\frac{\partial M}{\partial r}$= 0 can be used to calculate the degenerate horizon, yielding $$r_{dg} = 3r_+{}^2\Lambda$$Ò
The black hole's (\ref{sol}) entropy is computed using Eq. (\ref{ent}) and get:
\begin{eqnarray} \label{ent1}
{S_2}=-{\frac {\,c_4{r_2}^{15/2}{\Lambda}^{3/2}\pi}{12\sqrt {\Lambda
\,{r_2}^{3}-c_3\,{e^{{\frac {c_3}{\Lambda\,{r_2}^{3}}}}}}}
}.
\end{eqnarray}
Figure \ref{fig:2a} illustrates the entropy's behavior, displaying a positive value.

The black hole solution's (\ref{sol}) Hawking temperature is calculated as
\begin{eqnarray} \label{m44}
{T_2}=  {\frac {\left( 2\,{\Lambda}^{4}{r_2}^{9}+M \right)}{ 4{\pi}\left({\Lambda}^{4}{r_2
}^{9}-M \right)\sqrt {r_2{}^3\Lambda(\Lambda\,{r_2}^{3}-c_3\,{e^{^{^{\frac {c_3}{\Lambda\,{r_2}^{3}}}}})}}}},
\end{eqnarray}
In this situation, we call the Hawking temperature of the event horizon ${T_2}$. The temperature is shown in Figure 2, and it's always positive. Now, we'll calculate the Gibbs energy of  solution (\ref{sol}). Using Eq. (\ref{gib}), we get:
\begin{align} \label{enr}
&G(r_2)=\frac{{\Lambda}^{4}{r_2}^{9}\left( 48\,{\Lambda}^{17}{r_2}^{36}-48\,{
\Lambda}^{13}{r_2}^{27}c_3\,{e^{^{\frac {c-3}{{\Lambda}^{
4}{r_2}^{9}}}}}-48\,{\Lambda}^{4}{r_2}^{9}M+48\,Mc_3\,{e^{^{
\frac {c_3}{{\Lambda}^{4}{r_2}^{9}}}}}+2\,{r_2}^{36}{\Lambda}^{16}c_4+{r_2}^{9}{\Lambda}^{3}c_4\,M \right)}{48
\left( {\Lambda}^{13}
{r_2}^{27}-M \right) \left( {\Lambda}^{4}{r_2}^{9}-c_3\,{
e^{^{\frac {c_3}{{\Lambda}^{4}{r_2}^{9}}}}} \right) }.
\end{align}
 In the four-dimensional scenario, we graph the Gibbs function as shown in Figure 3, in cases where its value is positive. This indicates that global stability is reached by the black hole solution (\ref{sol}).

 \section{Discussion}\label{S12}

  The main discovery of this research is the general expression for $f({\cal Q})$ in a four-dimensional cylindrical spacetime. We've identified two unknown functions to describe it.  This form relies on a constant $c_3$. The analytic function $f({\cal Q})$ has a constant value when $c_3$ equals zero. This value corresponds to the linear case, that is, $f({\cal Q})={\cal Q}$. Also, the two unknown functions in the metric ansatz vary because of this constant, and they only match when $c_3=0$.  It is imperative to emphasis that, in contrast to what has happened in the literature \cite{Nashed:2023tua,sym16020219}, here and without making any assumptions, $f({\cal Q})$ is derived in its general form. We compute the invariants derived from both curvature and non-metricity in order to delve deeply into the black hole solution. We show that, for the radial coordinate, the singularity occurred at $r=0$. Our computations of the Kretschmann invariant show that it behaves as $K\sim r^{-1}$, in contrast to the Einstein-Maxwell theory solutions in the linear case of $f({\cal Q})$, which show $K\sim r^{-6}$. For the Kretschmann invariant, our black hole solution has a mild singularity compared to the linear case of $f({\cal Q})$.

Additionally, In order to create a stable and accurate four-dimensional rotating black hole, we used a coordinate transformation between temporal and angular angles. The rotational parameters are represented by $\xi_i$. The rotating black hole $f({\cal Q})$ assumes a non-trivial value, and is characterized by a dynamic scalar value ${\cal Q}$. In the end, we calculated the thermodynamic quantities related to the static black hole solution and proved that this black hole satisfies the thermodynamic requirements, meaning that its temperature, entropy, heat capacity, and Gibb's free energy are all positive.  The positive values of Gibb's energy, in particular, suggest that this black hole has thermodynamic stability, and are in line with the results reported in the literature cite{Nashed:2019tuk}.

The present research shied a light on the $f({\cal Q})$ theory that one can in general constitute its general form using certain spacetime, cylindrical spacetime, can we do this for other symmetry? This will be answer in elsewhere.


\end{document}